# Lead Nanopowder as Advanced Semi-Conductor – An Insight

**Thirugnanasambandan Theivasanthi[1*] and Marimuthu Alagar[1]**

[1]Centre for Research and Post Graduate Department of Physics, Ayya Nadar Janaki Ammal College, Sivakasi - 626124, Tamilnadu, India.

***Corresponding author:** Phone: +91-9245175532 *E-mail*: theivasanthi@pacrpoly.org

---

**Abstract:** *This work reports aspect related to semi-conducting nature of nano-sized particles of lead. This attempt finds its semiconducting behaviors elaborately and such advanced insight has so far not been said in literatures. New findings of Pb nanopowder (metal) by optical, fluorescence, cyclic voltammetry and four probe studies are presented in this study. UV-Vis analysis confirms the nano nature of the sample i.e. 5 nm and SPR peak confirms its metallic nature. The successful calculations of band gap, temperature dependent resistance value confirm its semi-conducting nature and explore its potential application in various industries. This work throws some light on and helps further research.*

---

## 1. Introduction

It is a general and well known fact that nano-materials are behaving differently from their bulk material. Size and shape control many of the physical properties (viz., melting point, magnetism, specific heat, conductivity, band gap, etc.), luminescence, optical, chemical and catalytic properties of nanomaterials. The present research has been done, based on these facts and the new nano sized Pb metallic semiconductor has been innovated. Photoluminescence study of lead nanopowder confirms the emission of photon. It suggests presence of bandgap in the material and its semiconducting properties.

A semiconductor is a material which has electrical conductivity between that of a conductor such as copper and an insulator such as glass. The conductivity of a semiconductor increases with increasing temperature, behavior opposite to that of a metal [1]. Most common semiconducting materials are crystalline solids. They are defined by their unique electric conductive behavior. They can pass current more easily in one direction than the other. Their conductive properties can be modified by controlled addition of impurities or by the application of electrical fields or light. Metal conductivity decreases with temperature increase because thermal vibrations of crystal lattice disrupt the free motion of electrons. Bulk lead metal is a good conductor but lead nanoparticles behaving differently and showing some resistance in four probe test. Semiconductor resistivity is indirectly proportional to temperature. It increases with decreasing temperature and vice versa. 4 eV bandgap value is the rough dividing line and it determines whether a material is semiconductor (less than 4 eV) or an insulator (more than 4 eV).

Youjun He *et al*. in their report, the absorption property, especially the absorption in visible region, are very important for the photovoltaic (PV) materials [2]. Low bandgap materials ($E_g$ < 1.8 eV) are of interest because their absorption spectra cover from the visible to the near-infrared region [3-4]. Semiconductors with band gap <1.8eV, or near-infrared light, have the greatest potential to form an efficient solar cell.

Surface Plasmon Resonance (SPR) of nano-metallic surfaces can absorb and intensify light at specific wavelengths. This is because the incoming light results in a collective oscillation of the electrons at the metal's surface. This plasmonics phenomenon has many promising applications and can be exploited to transmit optical signals, to interact with biomolecules and in solar cells. Metallic nanostructures can boost the absorption of light into photoactive materials of various types of solar cells like crystalline Si cells, cells based on high-performance III-V semiconductors, organic and dye-sensitized solar cells.

We have made an attempt to find semiconducting behaviors of Pb nanoparticles. In this study, we will present those new findings by optical, fluorescence, cyclic voltammetry and four probe studies. These findings suggest that the synthesized material is an efficient semiconducting material and can be utilized for making solar cells, optoelectronic, power and other semiconductor devices. To our knowledge, such advanced insights have so far not been said for Pb nanoparticles.

## 2. Experimental Details

In order to explore the semiconducting behaviors of lead nanopowder, we have synthesized it in accordance with our (T.Theivasanthi and M.Alagar) earlier literature procedure [5]. HRTEM analysis of the earlier report confirms the FCC structure of Pb nanoparticles and 10 nm average size of the spherical shaped particle. X-Ray Diffraction (XRD), Energy Dispersive X-Ray (EDAX), Differential Scanning Calorimetry (DSC), Atomic Absorption Spectroscopy (AAS), Fourier Transform-Infra Red (FT-IR) and theoretical density calculation studies of the same report confirms the sample was Pb metal Nanoparticles. UV-Vis analysis of the present study confirms the nano nature of the sample i.e. 5 nm and SPR peak confirms it's metallic nature.

5 g of $Pb(NO_3)_2$ salt was dissolved in 100 mL of distilled water and transferred to an electrochemical bath (volume: $4 \times 3 \times 3$ $cm^3$) with two electrode system - lead rod (anode) and stainless steel rod (cathode). A constant voltage of 15 V was applied between the electrodes using a power supply for a time span of 10 minutes. At the end of the process, deposition of Pb nanoparticles was observed and they were removed from the cathode and the Pb nanoparticles (settled down on the electrolytic cell) were also removed from the bath. 20 g of konjac tuber was sliced / cut into many pieces and boiled for 10 minutes with 100 mL of distilled water. At the end, konjac aqueous extract was decanted, few drops of extract was added to the synthesized Pb nanoparticles (to prevent oxidization and stabilization) and were kept in a hot air oven at 50 °C for two hours/until it dried.

UV-Vis analysis of the prepared nanopowder was done after dissolving in acetic acid, from wavelength 200nm to 1000nm, using VISIONpro software. Photoluminescence analysis of the sample was studied from wavelength 200nm to 900nm. Cyclic voltammetry was analyzed with a scan rate 0.1 V/s, using platinum electrode, silver reference electrode, platinum wire counter electrode and KCl electrolyte. A round shaped pellet of the sample was made and four probe analysis was done using this pellet.

## 3. Results and Discussions

### 3.1. Band gap

Zhenlin Wang *et al*. reported about the band gap of common metals like silver, copper and nickel [6]. The band gap is called "direct" if the momentum of electrons and holes is the same in both the conduction band and the valence band; an electron can directly emit a photon. If a material has a direct bandgap in the range of visible light, the light shining on it is absorbed, causing electrons to become excited to a higher energy state. Based on these facts and from photoluminescence study, direct bandgap of Pb Nanoparticles is assessed.

### 3.2. UV-Vis Analyses

Surface plasmons are created on the boundary of a metal. These represent quantized oscillations of surface charge produced by an external electric field [7]. The SPR excited on metal nanoparticles creates interesting optical properties [8]. In metal nanoparticles like gold, silver and copper, occurrence of dipole resonance in the UV-Vis region, making them useful for optical applications [9]. As well as the dielectric properties of the material, the frequency of the dipole resonance depends upon the size and shape of nanoparticles [10].

UV-Vis spectrum (absorbance Vs wavelength) of Fig.1 (a) shows SPR peak (Surface Plasmon Resonance) at 376 nm. The absorption, which corresponds to electron excitation from the valence band to conduction band, can be used to determine the nature and value of the optical band gap. SPR peak is an important and special characteristic feature of metal nanoparticles. It's presence in the UV spectrum of the sample confirms metallic nature i.e. the prepared sample is Pb metal nanoparticles. The calculated bandgap value confirms their semi-conducting nature.

Electron-pairing is universally agreed fact by which superconductivity occurs. Energy gap ($E_g$) is defined as the energy required for breaking up a pair of electrons and the amount of energy required for disrupting the superconducting state. The energy gap (in meV) of bulk lead is calculated as per BCS theory and it is described by

$$E_g = 7/2\,K\,T_c \qquad (1)$$

Where K is Boltzmann's constant ($8.62E^{-5}$ eV/K), $T_c$ is critical transition temperature in Kelvin. Bulk lead Energy gap is $3.381E^{-22}$ and $T_c$ is 7K.

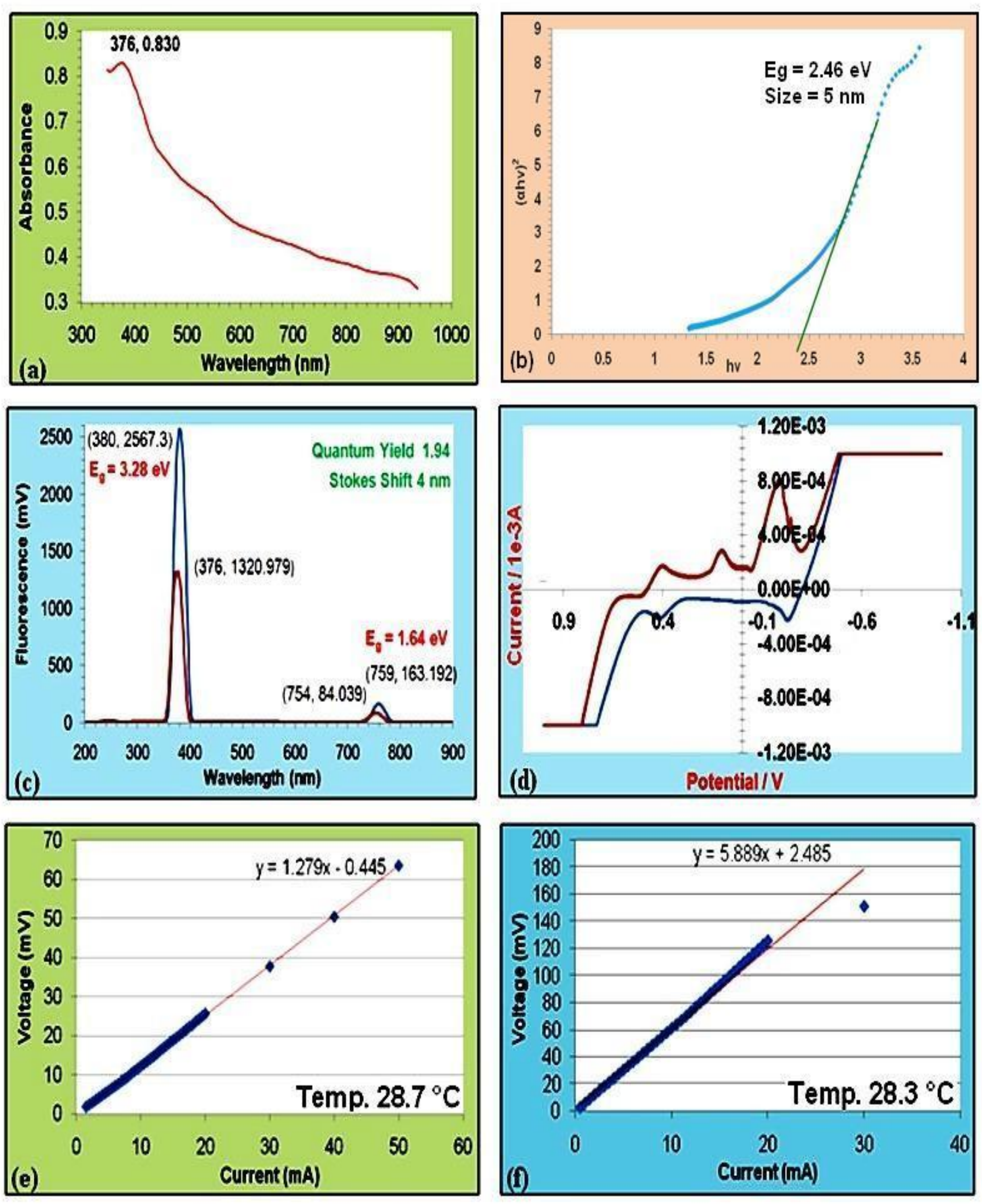

**Figure.1.** Various Analyses to confirm semi-conducting behavior of Pb nanopowder. (a) UV.Vis showing SPR peak at 376 nm. (b) Tauc plot. (c) PL spectrum showing Stokes shift. (d) CV showing oxidation & reduction peaks. (e&f) Four probe test at two different temperatures.

Sharma *et al*. in their report using Tauc relation to calculate, the relation between the absorption coefficient and the incident photon energy [11]. It is a convenient way of studying the optical absorption spectrum of a material. It is a method to determine the optical bandgap energy in semi-conductors. Tauc plot of the sample in fig.1b shows the photon energy (hν) on the X-axis, quantity $(\alpha h\nu)^2$ on the Y-axis and extrapolating the linear portion of the curve to the X-axis yields the energy of the material.

$$\alpha h\nu = A(h\nu - E_g)^n \quad (2)$$

$$\Delta E = nanoE_g - bulkE_g = h^2\pi^2/2MR^2 \quad (3)$$

Where, 'α' is absorption coefficient, hν is the incident photon energy, 'A' is constant which is different for different transitions and '$E_g$' and 'ΔE' is band gap of the material, exponent '$n = ½$' for direct transition, $nanoE_g$ is band gap of nano lead, $bulkE_g$ is band gap of bulk lead, h is Planck's constant, R is the radius of the particles and M is the effective mass of the system. Tauc plot of Fig.1 (b) determines the optical gap as 2.46 eV. The particle size of the prepared sample 5nm has been assessed from the equation (3) and it is well agreement with HRTEM particle size assessment of the same sample of our earlier report [5]. The presence of bandgap indicates that the prepared lead nanoparticles exhibit strong quantum confinement and confirms their semi-conducting nature.

### 3.3. Specific Surface Area Analyses

Specific surface area (SSA) is a derived scientific value that can be used to determine the properties of a material and has a particular importance in case of adsorption, heterogeneous catalysis and reactions on surfaces.

$$SSA = \frac{SA_{part}}{V_{part} * density} \quad (4)$$

$$S = 6 * 10^3 / D_p\rho \quad (5)$$

Where Vpart is particle volume and SApart is Surface Area of particle [12]; S is the specific surface area, Dp is the size of the particles (spherical shaped), and ρ is the density of Pb 11.461 $gcm^{-3}$ [13]. Both of these formulas can be used for mathematical calculation of SSA and both formulas will yield same result. Table.1 enumerates the calculated SSA results.

**Table.1.** Specific Surface Area of Lead Nanoparticles

| Particle Size (nm) | Surface Area ($nm^2$) | Volume ($nm^3$) | SSA ($m^2g^{-1}$) | SA to Volume Ratio | Density ($g\ cm^{-3}$) |
|---|---|---|---|---|---|
| 5 | 78.53 | 65.44 | 104.71 | 1.2 | 11.461 |

## 3.4. Photoluminescence Analyses

Photoluminescence spectrum of the sample is shown in Fig.1 (c). The PL efficiency increases at the decrease of the size. In other words, the small metal nanoparticles "lighten" better than the large ones. The peak of PL spectra determines the bandgap [14]. Jameson *et al*. reported that in a pure substance existing in solution in a unique form: the fluorescence spectrum is invariant, remaining the same independent of the excitation wavelength; it lies at longer wavelengths than the absorption; it is, to a good approximation, a mirror image of the absorption band of least frequency. Also, in their report, the presence of appreciable Stokes shift is principally important for practical applications of fluorescence because it allows to separate (strong) excitation light from (weak) emitted fluorescence using appropriate optics [15]. The PL analysis of the sample gives the value of quantum yield (QY) 1.94, Stokes shift 4 nm (the difference between positions of the band maxima of absorption and fluorescence of the same electronic transition), band gap ($E_g$) value 3.28 eV, size 4 nm for emission peak at 380 nm and 1.64 eV, size 6 nm for peak at 759 nm. This indicates that $E_g$ value is indirectly proportional to wavelength and particle size. The calculated $E_g$ value is 2.46 eV which corroborates with the $E_g$ value (2.46 eV) of UV-Vis analysis.

As a direct bandgap material, the visible light shining on its surface will be well absorbed by this sample material. Also, high specific surface area of the sample will augment its light absorbance property. QY value greater than 1 of this material is the result of the gain of energy and it shows possible utilization for heat or photochemical reaction or photo-induced or radiation-induced chain reactions, in which a single photon may trigger a long chain of transformations. These characters will be very useful while applying this semiconductor material in PV cells.

## 3.5. Cyclic voltammetry Analyses

Cyclic voltammetry is an effective method to characterize $E_g$ value of semiconductor materials. The $E_g$ value is very important value to be determined for photovoltaic materials. Al-Ibrahim *et al.* in a report, CV is one of the most accurate methods to characterize the organic materials and estimation about energy band diagram [16]. H.Karami *et al*. studied the relationship between particle size and electrochemical parameters. They concluded that smaller particles increase the currents of reduction and oxidation peaks [17]. CV study of lead nanoparticles does not coincide with H.Karami *et al*. conclusion. Reduction and oxidation peaks currents of our sample are less than the reduction/oxidation peaks currents earlier reports (electrochemical behavior of lead) by R.Salghi *et al*. [18] and W.Visscher [19].

CV of the sample has been conducted with a scan rate 0.1 V/s and shown in Fig.1 (d). Band gap of sample has been calculated from CV analysis as per report [20]. Band gap is determined by

$$E_g = I_p - E_a \quad \text{.......................................................(6)}$$

Where, $I_p$ is ionisation potential of material and $E_a$ is electron affinity. $I_p$ is calculated using oxidation onset (0.471) and $E_a$ using reduction onset (-0.259). The calculated $E_g$ value 0.73 eV has been compared with $E_g$ value from Tauc plot and PL analysis. The details are in Table.2.

### 3.6. Four probe analyses

Four probe analyses results of synthesized Pb nanoparticles has been shown Fig.1 (e & f). Resistance value (R) of this test indicates that the nano Pb acting differently from bulk Pb. R values of the sample are indirectly proportional to temperature and the values obtained from two different temperatures are R =1.279 Ω at 28.7 °C and R = 5.889 Ω at 28.3 °C. It is observed from this test, the value of R depends on temperature i.e. a slight reduction in temperature leads to larger increase in R value. This type of temperature dependent R value confirms the semi-conducting nature of sample.

### 3.7. Semi-Conductor Mechanism in Pb Nanoparticles

Metallic nanoparticles possess unique optical, electronic, chemical and magnetic properties that are strikingly different from that of the individual atoms as well as their bulk counterparts. The optical properties of nanoparticles are closely related to size-induced changes in the electronic structure and directly reflect the size-dependent energy structure of the particles. Spectroscopic methods probe the energy differences between two states for allowed transitions. The size effect on the optical absorption spectra of metallic nanocrystals is probably best known for the noble metal nanoparticles [21].

Nanomaterials exhibit interesting size- dependent electrical, optical and magnetic properties which are different from both atomic and bulk- level. At the nanoscale, because of quantum mechanics effects, they behave differently as compared to bulk materials. Indeed the emerging properties are due to the quantity variation in number of atoms i.e. in nanoparticles, number of atoms is less than their bulk matter where millions of atoms together in. Also, because of compactness, there are split in energy levels. So, even in magnetic applications, there is more control. For example, as the size decreases, bandgap of material increases [22-23].

**Table.2.** Comparison of Band gap, Particle size, No. of Atoms data of Pb Nanoparticles

| Analysis | Eg (eV) | Size D (nm) | No. of Unit cell | No. of Atoms |
|---|---|---|---|---|
| UV | 2.46 | 5 | 539 | 2156 |
| | 3.28 | 4 | 276 | 1104 |
| PL | 1.64 | 6 | 932 | 3728 |
| CV | 0.73 | 9 | 3186 | 12744 |

The number of unit cell for the synthesized Pb nanoparticles is calculated from

$$n = \pi(4/3)\ \text{x}\ (D/2)^3\ \text{x}\ (1/V) \quad \ldots\ldots\ldots\ldots(7)$$

Where D is the crystallite size and V is the cell volume of the sample [24]. The calculated details are enumerated in Table.2. Only a small quantity (between the range 1000 to13000), number of atoms is presence in a particle which is very much lower than bulk Pb. It clearly indicates the quantum mechanics effects of the sample. This effect plays an important role in splitting / discrete / enhancing the energy levels (bandgap) of the sample. Figure.2 shows increasing of the bandgap while decreasing number of atoms.

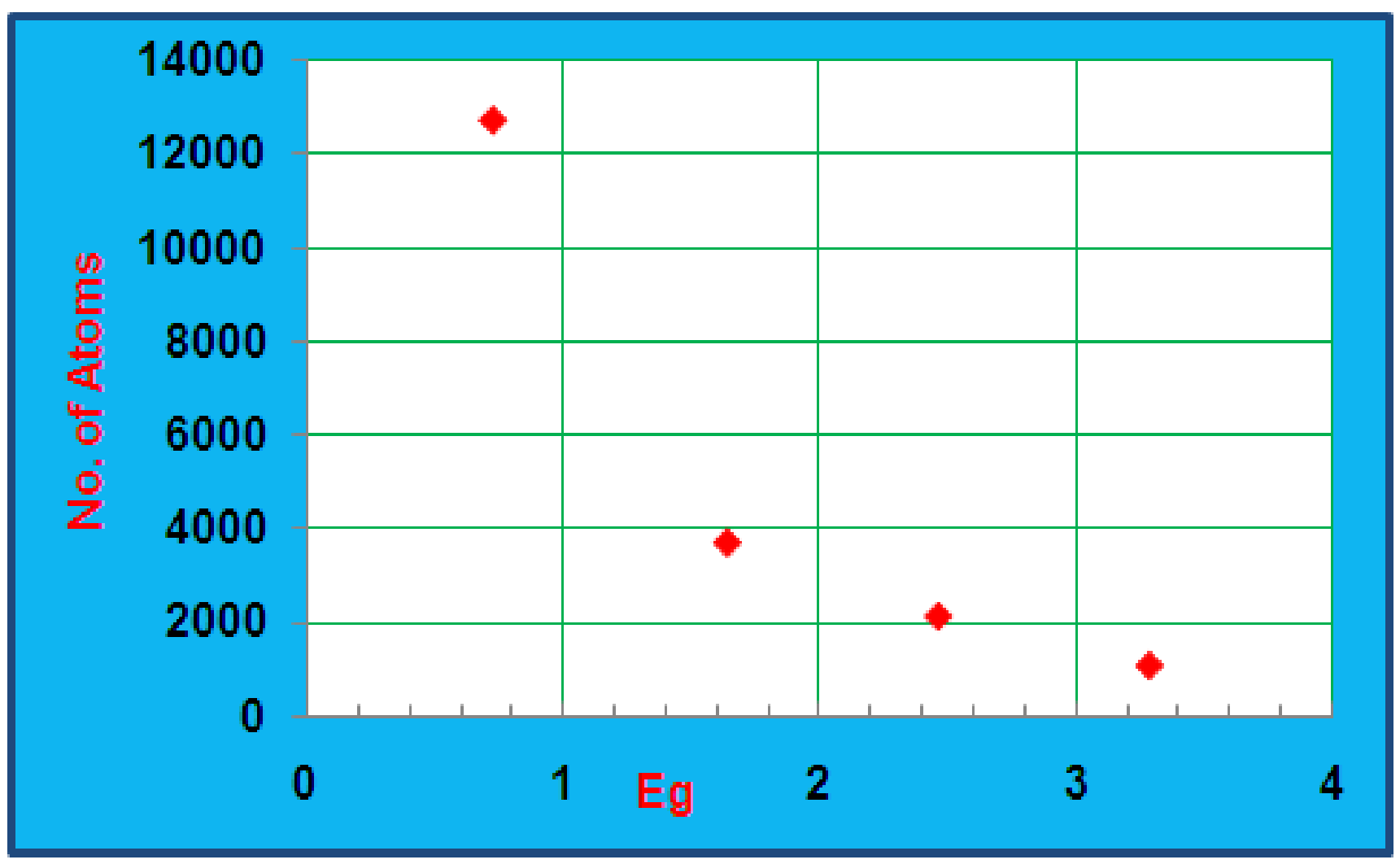

**Figure.2.** Bandgap Vs Number of Atoms in Pb Nanoparticles

Small metal nanoparticles exhibit the absorption of visible electromagnetic waves by the collective oscillation of conduction electrons (stimulated / excited by incident light) at the surface [25]. This is known as the Localized Surface Plasmon Resonance effect. This effect is useful for tracing the presence of metal nanoparticles with a simple UV-visible spectrometer. The size dependence of the SPR for particles smaller than 20 nm is a complex phenomenon. The bandwidth of resonance increases while size of the particles decreases due to electron scattering enhancement at the surface. This is one of the interesting parameters to characterize metal nanoparticles [26].

At nanometer scale, an object has a high percentage of their constituent atoms at the surface [27]. The volume decreases more quickly than its surface area, as the size diminishes; in the most extreme case, this scaling behavior leads, to the structure where every atom in the structure is interfacial and in some sense, it could be assumed that nanostructures are “all surface” [28]. Quantum confinement effects cause changes in the spacing between intraband energy levels of metallic nanoclusters i.e. energy levels increases with decreasing metallic nanocluster size. solutions of noble metals *viz.* copper, silver, gold show a very intense colour and exhibit a strong absorption band, which is absent in the bulk material as well as for the individual atoms. This absorption band results the localised surface Plasmon resonance [29]. The strict criterion for metallicity, a finite density of states at the Fermi energy, cannot be applied to clusters because energy levels are always discrete in a system of finite size [30].

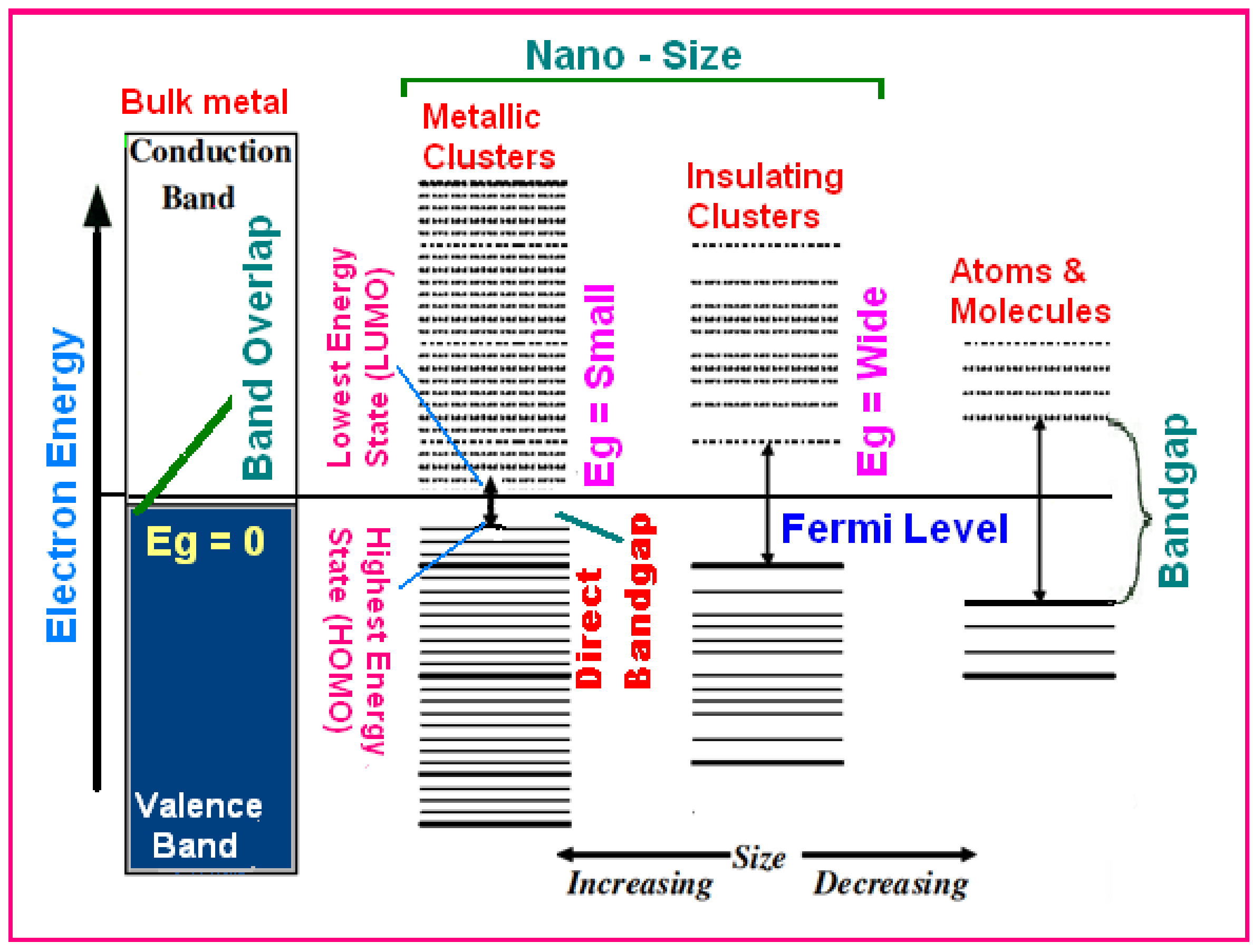

**Figure.3.** Schematic presentation - Band gap Vs density of states (number of atoms)

A transition from the bulk band structure to individual localized energy levels occurs in clusters of subnanometer to nanometer size, and the detection of quantum size effects has been of great interest to scientists and technologists in the search for novel materials with exciting new properties [25]. The overall properties of the new material are determined by the properties of the individual particles as a function of their size and shape as well as their collective behavior [31].

As the dimension of the particles shrinks into the nanometer regime, there are significant changes in the physicochemical properties due to both quantum size effects and the important role of the increased surface in controlling the overall energy of the particles. The nano-scale generates novel properties that can hardly be seen in the bulk, such as, the conductor-insulator and nonmagnetic-magnetic transition of noble metals and tuning of constant physical quantities such as, Young's modulus, dielectric constant, melting point etc. Moreover, at 2-3 nm range, gold nanoparticles cease to be noble and are excellent catalysts. At this size they are still metallic, but turn into insulators [32].

When all three dimensions of a piece of material in nano-scale, it is referred to as quantum dot. Semiconductor quantum dots in a matrix have energy bandgap higher than the quantum dot. Hence, they can trap charge carriers. They exhibit enhanced energy gap, discretization of the electronic levels, large oscillator strength, high excitonic Bohr radius and very high surface-to-volume ratio [33]. In band theory, the breadth of a band is directly related to the strength of interactions among nearest neighbors [34].

In condensed matter physics a band gap means an energy range in a solid, where no electron states are allowed. This bandgap controls electrical and optical properties. The changes in bandgap, lead to the changes in the electrical and optical properties of the solid. A direct bandgap means that the minimum (lowest energy state) of the conduction band lies directly above the maximum (highest energy state) of the valence band in momentum space. This makes it more likely for an electron to be excited and therefore makes the semiconductor more sensitive. In a direct bandgap semiconductor, electrons at the conduction-band minimum can combine directly with holes at the valence band maximum, while conserving momentum. The energy of the recombination across the bandgap will be emitted in the form of a photon of light. This is radiative recombination, also called spontaneous emission.

Electrical band gap is simply the difference between the top of the valence band (HOMO - Highest Occupied Molecular Orbital), and the bottom of the conduction band (LUMO - Lowest Unoccupied Molecular Orbital). It refers to the differences of oxidation (add a hole or remove an electron from the HOMO) and reduction (add an electron to the LUMO) of molecular species. In a photo-excitation, optical bandgap refers to a strongly bound electron-hole pair by transferring an electron from HOMO to LUMO (or from the ground state to an excited state). In the case of inorganic materials, direct optical band gap is always greater than the electrical band gap because the energy of an unoccupied state is changed upon occupation.

Bose et al. have observed a systematic crossover from metallic to weakly insulating behavior with a reduction in the average grain size in nanostructure films of elemental niobium. It has been shown that the energy barrier at the grain boundary increases with a decreasing grain size. When, the critical grain size is 8 nm, the metal to insulator transition occurs in nanostructure niobium due to the opening up of an energy gap at the grain boundary [35].

Discreteness of energy levels also brings about changes in the spectral features, especially, those related to the valence band [36]. The electrical conductivity and magnetic susceptibility are both governed by spacing between the adjacent energy levels and exhibit quantum size effects [37]. The electronic structure of the nanoparticles depends critically on the particles size [38]. Very few atoms comprise in discrete nanoclusters and the spacing between adjacent energy levels result a shift in conductive properties of the nanocluster, from metallic to semiconducting and insulating with decreasing size [39]. Fig.3. shows shifting of conductive properties from metallic to semiconducting / insulating in discrete nanoclusters.

In summary, optical properties of sample lead metal nanoparticles directly reflects the size-dependent energy structure of the particles and the UV absorption spectrum is the best one to probe the energy differences between two states of valence and conduction bands; the sample exhibits electrical (electronic – semiconductor) properties different from bulk lead, at room temperature due to quantum mechanics effects; possess very small number of atoms; Energy levels ($E_g$) have been splitted / enhanced (bandgap increases while size decreases due to electron scattering enhancement at the surface) but in case of bulk lead $E_g = 0$; bandgap enhancement (the interesting parameter to characterize metal nanoparticles) in the sample confirms lead metal nanoparticles; the volume is less than surface area in Pb nanoparticle; it leads to all the atoms in the particle is interfacial and the all particles in the sample are - all surface; a high percentage of Pb atoms accumulate at the surface which exhibits the special features of Pb metallic nanoclusters like strong absorption band and localized Surface Plasmon Resonance (SPR); these features are caused by the quantum confinement effects of Pb metallic nanoclusters as said earlier and could not be seen in bulk Pb or in Pb atom or in lead oxides;

SPR is a parameter in the characterization of Pb metal nanoparticles and it is useful for tracing Pb nanoparticles; transition from the bulk band structure to individual localized energy levels of Pb nanoclusters is also observed and the overall properties of this new material like SPR is the collective behavior of all Pb particles; strict criterion for metallicity like good conductivity and overlapped conduction-valence bands at the Fermi energy are not applicable to such finite sized Pb clusters system because energy levels are always in discrete state; The nano-scale generates novel properties like transition of conductor-insulator and nonmagnetic-magnetic transition have already been reported in various literatures; 2-3 nm range nano metallic gold and nanostructured films of elemental niobium turn into insulator, likewise the sample nano metallic Pb turns into semiconductor; the quantum dot structure of the sample enhances energy gap and discretes the electronic levels which leads to the shift in conductive properties of the nanocluster, from metallic to semiconducting; photoluminescence activities of the sample emit photon of light which clearly indicates the direct bandgap of the sample; this optical bandgap in the sample is higher than electrical bandgap observed from CV; resistance value of the sample decreases while temperature increases (i.e.conductivity increases while temperature increases) – a semiconductor behavior - opposite to metal behavior is observed in Four probe analyses.

## 4. Conclusions

First time, we have made an attempt to find semiconducting behaviors of Pb nanoparticles (metal) and such advanced insight have so far not been said in literatures. We have calculated its band gap value and measured its temperature dependent resistance (R) value successfully. Band gap value, below the rough dividing line (less than 4 eV) and temperature dependent resistance (R) value confirms its semi-conducting nature. The results suggest that as a Semiconducting material with a high QY i.e. more than 1, it will form an efficient solar (PV) cell. SPR of this nano-metallic surface can boost the absorption of and intensify light into PV Cells. Also, it has a lot of potential applications in the industries of optoelectronic, power & other semiconductor devices. This work throws some light on and helps further research on nano-sized lead powder.

**Acknowledgements**

The authors express thanks to Lady Doak College, Madurai, Thiagarajar College of Engineering, Madurai and Mepco Schlenk Engineering College, Sivakasi (India) for providing analytical instruments facility.